\input phyzzx \def\e{\adveq\eqno{\rm (\chapterlabel.\the\equanumber)}}
\def\mysec#1{\equanumber=0\chapter{#1}}
\def\adveq{\global\advance\equanumber by 1}
\def\myeq{{\rm \chapterlabel.\the\equanumber}}
\def\rarrow{\rightarrow}

\def\twoline#1#2{\displaylines{\qquad#1\hfill(\adveq\myeq)\cr\hfill#2
\qquad\cr}}

\def\semidirect{\mathrel{\raise0.04cm\hbox{${\scriptscriptstyle |\!}$
\hskip-0.175cm}\times}}

\def\mod{\mathop{\rm mod}\nolimits}

\def\ref#1{$^{[#1]}$}

\def\r#1{$[\rm#1]$} 
\def\twidle{\tilde}

\def\threeline#1#2#3{\displaylines{\qquad#1\hfill\cr\hfill#2\hfill\llap{(\adveq\myeq)}\cr
\hfill#3\qquad\cr}}

\def\e{\adveq\eqno{\rm (\chapterlabel.\the\equanumber)}}
\def\mysec#1{\equanumber=0\chapter{#1}}
\def\adveq{\global\advance\equanumber by 1}
\def\myeq{{\rm \chapterlabel.\the\equanumber}}
\def\rarrow{\rightarrow}

\def\twoline#1#2{\displaylines{\qquad#1\hfill(\adveq\myeq)\cr\hfill#2
\qquad\cr}}

\def\semidirect{\mathrel{\raise0.04cm\hbox{${\scriptscriptstyle |\!}$
\hskip-0.175cm}\times}}

\def\mod{\mathop{\rm mod}\nolimits}

\def\ref#1{$^{[#1]}$}

\def\r#1{$[\rm#1]$} 
\def\twidle{\tilde}

\def\threeline#1#2#3{\displaylines{\qquad#1\hfill\cr\hfill#2\hfill\llap{(\adveq\myeq)}\cr
\hfill#3\qquad\cr}}

\overfullrule=0pt

\date{July,  2020}
\date{July, 2020}
\titlepage
\title{On SO$(N)$ Spin Vertex Models }
\author{Vladimir  Belavin$^{a}$, Doron Gepner$^b$, Hans Wenzl$^c$}
\vskip20pt
\line{\it\hfill  $^a$\  Physics Department, Ariel University, Ariel 40700, Israel \hfill}
\line{\it\hfill $^b$ Department of Particle Physics and Astrophysics, Weizmann Institute,\hfill}
 \line{\it\hfill Rehovot 76100,  Israel\hfill} 
 \line{\it\hfill $^c$ Department of Mathematics, University of California, San Diego, California\hfill}

\abstract

We describe the Boltzmann weights of the $D_k$ algebra spin vertex models. Thus, we find  the
$SO(N)$ spin vertex models, for any $N$, completing the $B_k$ case found earlier. 
We further check that the real (self--dual) SO$(N)$ models obey
quantum algebras, which are the Birman--Murakami--Wenzl (BMW) algebra for three blocks,
and certain generalizations, which include the BMW algebra as a sub--algebra, for four and five blocks. In the case of five blocks, the $B_4$ model
is shown to satisfy additional twenty new relations, which are given. The $D_6$ model is shown to obey
two additional relations.

\endpage 

\mysec{Introduction.}

Solvable lattice models in two dimensions are a fruitful ground to test phase transitions, universality,
integrability 
\REF\Zam{A.B. Zamolodchikov and A.B. Zamolodchikov, Ann. Phys. (NY) 120 253 (1979).}
\r\Zam\
and conformal field theory
\REF\Francesco{P. Francesco, P. Mathieu and D. Senechal, ``Conformal field theory", Springer (1997).}
\r\Francesco.
For reviews see
\REF\Baxter{R.J. Baxter, ``Exactly solved models in statistical mechanics", Academic Press,
London, England (1982).}
\REF\Wadati{M. Wadati, T. Deguchi and Y. Akutsu, Phys. Rep. 180 (4) (1989) 247.}
\r{\Baxter,\Wadati}. 

We will concentrate here on a type of solvable lattice models which are called
vertex models. Well known among these are the six, eight and nineteen vertex models
\r{\Baxter,\Wadati}.
For recent works on vertex models see, e.g.,
\REF\Bru{B. Brubaker, arXiv: 1906.04140 (2019).}
\REF\Kas{T.K. Kassenova, P. Yu and O.V. Razina, J. of Phys.: Conf. Series
1391 (2019).}
\REF\Bos{A. Bossard and W. Galleas, J. of Math. Phys. 60, 103509 (2019).}
\REF\Arb{J.H. Arbeitman, S. Mantilla and I. Sodeman, Phys. Rev. B 99 245108 (2019).}
\REF\Nir{K.S. Nirov and A.V. Razumov, SIGMA 15 068 (2019).} 
\r{\Bru,\Kas,\Bos,\Arb,\Nir}.
Our purpose here is to introduce vertex models based on the algebra $D_k$ and the spin representation.
This completes the SO$(N)$ spin vertex models for all $N$, where the $B_k$ models were 
described before in
\REF\Bk{D. Gepner, ``$B_k$ spin models and quantum algebras", arXiv: 2005.02708, Nucl. Phys. B
(in press) (2020) .}
ref. \r\Bk.

We are also interested in the algebraic structure underlying these models. We use 
the more general results of
\REF\CBfour{V. Belavin, D. Gepner, J.R. Li and R. Tessler, JHEP 11 (2019) 155.}
\REF\CB{V. Belavin and D. Gepner, arXiv: 2001.09280 (2020).}
\r{\CBfour,\CB}, which describe the three, four and five blocks algebras
(where the number of blocks is the degree of polynomial equation obeyed by the Boltzmann weights),
assuming only a certain ansatz for the Baxterization, described in
\REF\Found{D. Gepner, ``Foundations of rational quantum field theory I", arXiv: hep-th/9211100v2 (1992).}\r\Found,
and the Yang--Baxter equation.
We describe and check numerically, the algebras of $B_4$, which is a five blocks theory,
and the algebra of $D_6$, which is a four blocks theory. 

The algebras include
a version of the Birman--Murakami--Wenzl algebra (BMW)
\REF\BW{J.S. Birman and H. Wenzl, Trans. Am. Math. Soc. 313 (1) (1989) 313.}
\REF\Mur{J. Murakami, Osaka J. Math. 24 (4) (1987) 745.}
\r{\BW,\Mur}, along with two new relations for four blocks and twenty new relations for the five blocks theory, which are given here for $B_4$. We check that the BMW algebra is obeyed for $D_k$, for any small even 
$k$, with a different skein relation.

\mysec{$D_k$ spin vertex models.}

We wish to describe a vertex model based on the algebra $D_k=SO(2k)$ and the spin representation.
This solution is an element of End$(V\otimes V)$ where $V$ is the spin representation of
$D_k$. We denote by $\alpha_n=\epsilon_n-\epsilon_{n+1}$, for $n=1,2,\ldots,k-1$ and
$\alpha_k=\epsilon_{k-1}+\epsilon_k$ the simple roots of $D_k$, where $\epsilon_i$ are 
orthogonal unit vectors. The spin representation, denoted by $S$ has the weights
$\sum_{i=1}^k p_i \epsilon_i/2$, where $p_i=\pm 1$ and $\prod_{i=1}^k p_i=1$.
The last product is $-1$ for the anti--spinor representation, denoted by $\bar S$.
We find it useful to add $1/2$ to these weights, and to represent weights of the spinor (anti--spinor) 
representation by the vector $m$, where $m_i=0$ or $1$.

To start constructing the vertex model, we need a solution which commutes with the co--product
of $U_{q^2}(SO(2k))$. We find it convenient to first describe a solution for the larger representation
$\twidle V=S\oplus \bar S$, namely the sum of the spinor and anti--spinor representations.
Such a solution was described recently in
\REF\Wenzl{H. Wenzl, ``Dualities for spin representation", arXiv: 2005.11299, (2020).}\r\Wenzl. It is the element $C$ of End$(\twidle V \otimes \twidle V)$,
given by
$$C_{m,n}^{b,c}=\sum_{j=1}^k \delta_{m_j,1-n_j} (-q^2)^{\{m-n\}_j} \delta_{b,{\bar m}_j}
\delta_{c,{\bar n}_j},\e$$
where 
$$\{ m\}_j=\sum_{r=1}^j m_r,\e$$
and ${\bar n}_j$ is equal to $n_j$ except at the $j$th coordinate where it is $1-n_j$. Here 
$m,n,b,c=0$ or $1$ are weights of the spin or anti--spin representations shifted by
$1/2$. The eigenvalues of the matrix $C$ were computed in ref.  \r\Wenzl, and are
$$\lambda_j=\pm s(k-j),\qquad {\rm for\ } j=0,1,\ldots,k,\e$$
where
$$s(x)={q^{2 x}-q^{-2 x} \over q^2-q^{-2}}.\e$$

The solution $C$ has the disadvantage of mapping both the spin and anti--spin representations.
We note, however, that $C$ maps the representation $S\otimes S$ to $\bar S\otimes\bar S$,
and vice versa. Thus, to get a solution in End$(S\otimes S)$ all we need to do is to square the 
matrix $C$ and to equate to zero all the $C_{m,n}^{b,c}$ for weights $m,n,b,c$ which are not in $S$.
Thus, $C^2$ gives the solution we want. Of course, since $C$ commutes with the co--product,
so does $C^2$.
 
Since the matrix $C^2$ commutes with the co--product, it has the same eigenvectors as our desired 
solution which obeys the Yang--Baxter equation, but not the same eigenvalues. Thus, we define
the projection operators
$$ (P^a)_{m,n}^{b,c} =\prod_{p\neq a} \left[ {C^2-\lambda_p^2 I \over \lambda_a^2-\lambda_p^2 }\right],\e$$
where the product is in End$(S\otimes S)$ and $I$ is the identity map.

We note that for even $k$, $P^a=0$ for $a$ which is odd, whereas for odd $k$, $P^a=0$ for
$a$ which is even. The $j$th eigenvalue corresponds to the representation
$V_j=\wedge^j v$, where $v$ is the vector representation, i.e., the anti--symmetric product
of $j$ vector representations \r\Wenzl. 
The highest weight of the representation $V_j$ is $\epsilon_1+\epsilon_2+\ldots +\epsilon_j$.
Thus, the non--zero $P^a$ are in one to one correspondence
with the representations that appear in the tensor product,
$$ S \times S=\sum_{j=0\atop j=k \mod 2}^k V_j,\e$$
as they should. Thus, the projection $P^a$ projects onto the representation $V_a$.

We wish to make the connection between the solution $C^2$ and the $D_k$ WZW conformal model.
For explanation of conformal field theory see the book
\r\Francesco, and references therein.
To do this we define,
$$q^2=\exp[\pi i/(r+g)],\e$$
Here $r$ is the level of the representation and 
$$g=2 k-2,\e$$
is the dual Coxeter number.

The dimension of the highest weight $\Lambda$  in a WZW theory is given by
$$\Delta_\Lambda={\Lambda(\Lambda+2 \rho)\over 2(r+g)},\e$$
where $\rho$ is half the sum of positive roots and $C_\Lambda=\Lambda(\Lambda+2 \rho)$ is
the Casimir of the representation. The Casimir of the representation $V_j$ is given by
$$C(V_j)=C_j=j(2 k-j).\e$$

As explained in
\r\Bk, the eigenvalues of the $R$ matrix are given by
$$\beta_j=p_j e^{-i\pi \Delta_j}=p_j q^{-C(V_j)},\e$$
where $p_j=\pm 1$ is some sign which corresponds to whether the product in eq. (2.6) is symmetric
or anti--symmetric. In our case, the sign is given by
$$p_j=(-1)^{(k-j)/2}.\e$$
Thus, since we know the eigenvalues of the $R$ matrix and the projection operators
from eq. (2.5), we may construct the $R$ matrix as
$$R^{a,b}_{m,n}=\sum_{j=0}^k \beta_j (P^j)^{a,b}_{m,n}.\e$$

It can be verified that this $R$ matrix satisfies the Yang--Baxter equation (YBE) which for the
$R$ matrix is the braiding relation,
$$\sum_{\alpha,\beta,\gamma} R_{j,k}^{\beta,\alpha} R_{i,\beta}^{l,\gamma}  R_{\gamma,\alpha}^{m,n} =
\sum_{\alpha,\beta,\gamma} R_{i,j}^{\alpha,\beta}  R_{\beta,k}^{\gamma,n}  R_{\alpha,\gamma}^{l,m}.\e$$
We checked that this $R$ matrix obeys the YBE, numerically for $k=2,3,4,5,6$ and it holds, indeed, for various
weights and for general $q$.

Now, we wish to define a trigonometric solution for the YBE. For this purpose, we use the same 
general ansatz
for Baxterization as in
\r{\Bk,\Found}. First, we need to decide on the order of the primary fields in eq. (2.6).
The order which solves the YBE is given by
$$(h_0,h_1,\ldots,h_{k/2})=(0,2,4,\ldots,k),\e$$
for even $k$. For odd $k$ the order is
$$(h_0,h_1,\ldots,h_{(k-1)/2})=(k,k-2,k-4,\ldots,1).\e$$
The parameters are given by \r{\Bk,\Found},
$$\hat \zeta_j=\pi(\Delta_{h_{j+1}}-\Delta_{h_j})/2,\e$$
for $j=0,1,\ldots,m-1$, where $m=k/2$ for even $k$ and $m=(k-1)/2$ for odd $k$.
Thus, the $D_k$ theory is an $m+1$ blocks theory.
We thus define the  parameters as
$$\zeta_j=(C_{h_{j+1}}-C_{h_j})/2.\e$$

We define
$$p(x)=q^x-q^{-x}.\e$$
Then the trigonometric solution to the YBE assumes the form \r{\Bk,\Found},
$$R_{m,n}^{a,b} (u)=\sum_{j=0}^{m} f_j(u) (P^{h_j})_{m,n}^{a,b} ,\e$$
where
$$f_a(u)=\left[ \prod_{j=1}^a p(\zeta_{j-1}-u) \right ] \left[ \prod_{j=a+1}^{m} p(\zeta_{j-1}+u)\right]\bigg/
\left[ \prod_{j=1}^{m} p(\zeta_{j-1})\right],\e$$
where $a=0,1,\ldots,m$.

For example, for $k=6$, which is a four blocks theory, the parameters are
$(\zeta_0,\zeta_1,\zeta_2)=(10,6,2)$. The crossing parameter is $\lambda=\zeta_0$.
The $D_k$ vertex models with $k$ even are real (self--dual) as $S= S^*$. For odd 
$k$ the theories are not real (not self--dual), as $S\neq S^*$.

We can check that the solution, eqs. (2.20, 2.21), obeys the Yang--Baxter equation, which is
$$\sum_{\alpha,\beta,\gamma} R_{j,k}^{\beta,\alpha}(u) R_{i,\beta}^{l,\gamma} (u+v) R_{\gamma,\alpha}^{m,n} (v)=
\sum_{\alpha,\beta,\gamma} R_{i,j}^{\alpha,\beta} (v) R_{\beta,k}^{\gamma,n} (u+v) R_{\alpha,\gamma}^{l,m}(u).\e$$
We checked this equation, numerically, for $k=2,3,4,5,6$ and various values of $u$, $v$ and $q$
and various heights. It is indeed obeyed. This gives the trigonometric $D_k$ spin vertex model.

\mysec{BMW$^\prime$ algebra and SO$(N)$ spin vertex models.}

We repeat here the definition of the BMW$^\prime$ algebra following  \r\Bk.
We find it convenient to use an operator form for the $R$ matrix. We define the matrix,
following 
\r\Wadati,
$$X_i(u)=\sum_{m,n,a,b} R^{a,b}_{m,n}(u) I^{(1)} \otimes \ldots \otimes I^{(i-1)} e^{(i)}_{am}
\otimes e_{bn}^{(i+1)}\otimes I^{(i+2)}\otimes\ldots\otimes  I^{(f)},\e$$
where $I^{(i)}$ is the identity matrix at position $i$ and $(e_{rs})_{lm}=\delta_{rl} \delta_{sm}$.
The YBE, eq. (2.22), then assumes a more compact form,
$$X_i(u) X_j(v)=X_j(v) X_i(u),\qquad{\rm if\ } |i-j|\geq2,$$
$$X_i(u) X_{i+1}(u+v) X_i(v)=X_{i+1}(v) X_i(u+v) X_{i+1}(u).\e$$ 

Let us denote the number of blocks by $n$. For the $D_k$ models, this is $n=m+1=k/2+1$ ($k$
even), or $n=m+1=(k+1)/2$ (for odd $k$). In this section, we will assume that $k$ is even,
so that the theory is real (self--dual). 
It is assumed that the number of blocks is greater or equal to three, $n\geq3$.
The algebras of non--real theories are also interesting,
but we shall not describe it here.
We define the limit of the matrix $X_i(u)$ as
$$X_i=\lim_{u\rarrow i\infty} e^{i(n-1) u} X_i(u),\qquad X_i^t=\lim_{u\rarrow-i\infty} 
e^{-i(n-1) u} X_i(u).\e$$

We define the operators,
$$G_i=2^{n-1} e^{-i(n-1)\zeta_0/2}\left[ \prod_{r=1}^{n-1}\sin(\zeta_{r-1})\right] X_i,\e$$
$$G_i^{-1}=2^{n-1} e^{i(n-1)\zeta_0/2} \left[\prod_{r=1}^{n-1} \sin(\zeta_{r-1})\right] X_i^t,\e$$
and
$$E_i=X_i(\zeta_0),\qquad 1_i=X_i(0),\e$$
where $\zeta_i$ are the parameters defined in eq. (2.17).
$G_i^{-1}$, so defined, is the inverse of $G_i$, or $G_i G_i^{-1}=1_i$.

From the ansatz, eqs. (2.20, 2.21),  and from the YBE, eq. (2.22), we can prove the following relations of the
operators $G_i$, $G_i^{-1}$ and $E_i$,
$$E_i E_{i+1} E_i=b E_i,\qquad E_i^2=b E_i,\qquad E_i E_j=E_j E_i \quad {\rm if \ } |i-j| \geq 2,\e$$
$$b=\prod_{r=1}^{n-1} {\sin(\zeta_0+\zeta_{r-1})\over \sin(\zeta_{r-1})},\e$$
which is the Temperley--Lieb algebra
\REF\TL{N. Temperley and E. Lieb, Proc. R. Soc. A 322 (1971) 251.}
\r\TL, and
$$G_i G_j=G_j G_i \quad {\rm if\ } |i-j|\geq2, \qquad G_i G_{i+1} G_i=G_{i+1} G_i G_{i+1},\e$$
which is the braiding algebra.
We can also prove the relations,
$$G_i E_i=E_i G_i=l^{-1} E_i,\e$$
where 
$$l=i^{n-1} \exp\left[ i(n-1)\zeta_0/2+i \sum_{r=0}^{n-2} \zeta_r\right].\e$$
The following is the skein relation which stems from the definition of the projection operators
along with the ansatz, eqs. (2.20, 2.21),
$$G_i^{n-2}=a E_i+\sum_{r=-1}^{n-3} b_r G_i^r,\e$$
where $a$ and $b_r$ are some coefficients, which can be expressed in terms of the parameters, 
$\zeta_r$.
From the skein relation we prove,
$$G_{i\pm 1} G_i E_{i\pm 1}=E_i G_{i\pm 1} G_i.\e$$

The above relation, eqs. (3.7--3.13), are part of the Birman--Murakami--Wenzl algebra (BMW)
\r{\BW,\Mur}.
The rest of the relations of the BMW algebra are also obeyed, except of the skein relation,
eq. (3.12), which is different for more than three blocks. These are
$$G_{i\pm1} G_i E_{i\pm1}=E_i E_{i\pm1},\qquad G_{i\pm1} E_i G_{i\pm1}=G_i^{-1} E_{i\pm1} G_i^{-1},\e$$
$$G_{i\pm1} E_i E_{i\pm1}=G_i^{-1} E_{i\pm 1},\qquad E_{i\pm1} E_i G_{i\pm1}=E_{i\pm 1} G_i^{-1},\e$$
$$E_i G_{i\pm 1} E_i=l E_i,\qquad E_i G_{i\pm1}^{-1} E_i=l^{-1} E_i.\e$$

We have verified that the full BMW$^\prime$ algebra (BMW with a different skein relation) is obeyed 
by the $D_k$ model, with $k=4$ or $6$. We did this numerically, using various heights and 
general $q$.
We note that the BMW$^\prime$ algebra is obeyed also by the $B_k$ spin vertex models
\r\Bk, while substituting the relevant parameters $\zeta_r$. For more than three blocks there are 
additional relations, except from the skein relation. These are described in the next two sections,
for $B_4$ and $D_6$.

\def\center#1\endcenter{\centerline{#1}}
\def\threeline#1#2#3{\displaylines{\qquad#1\hfill\llap{(\adveq\myeq)}\cr\hfill#2\hfill \cr
\hfill#3\qquad\cr}}

\def\frac#1#2{{#1\over #2}}

\mysec{$n=5$ blocks case and $B_4$ vertex models.}
It was noticed in \r\Bk\  that the structure of n-CB algebra, which follows from the Baxterization of IRF models, is also applicable for vertex models.
In our case, the connection is between n-CB  algebra and  $B_{n-1}$ models.  This algebra was checked, in particular, for $B_3$ models obeying 4-CB algebra \r\Bk. In the 5-block case, the n-CB algebra reduces to a set of $20$ relations in addition to BMW$^\prime$ sub--algebra.
We get these relations by expanding the YBE, eq. (2.22) and assuming the ansatz, eqs. (2.20, 2.21). 
The 5--CB relations are general for all the five blocks models obeying the ansatz for Baxterization, eq. (2.20, 2.21).
We specify the algebra here only for the $B_4$ spin vertex models for calculation reasons.
In this section we summarize the 5-CB relations for $B_4$ spin vertex model. We 
give the shorter relations explicitly, here. The complete  list of 5-CB relations for $B_4$ models can be found in the attached Mathematica file.

For the general values of the parameters, the 5-CB skein relation as well as the explicit projectors have been found in 
\r\CB.
The skein relation reads
$$\eqalign{
&G_i^3= \alpha  1_i+\beta  E_i+\gamma  G_i + \delta  G^{-1}_i +\mu G^2_i\;,
}
\e $$ 
where denoting $s_k = q^{\zeta_k}$ the parameters are
$$\eqalign{
& \alpha = -\frac{s_1 \left(s_1^2 s_2^2 s_3^2-s_2^2 s_3^2+s_3^2-1\right)}{s_0^3 s_2 s_3^3}\;,\cr
&\beta = \frac{\left(s_1^2-1\right)\left(s_2^2-1\right)  \left(s_0^2 s_1^2 s_2^2+1\right) \left(s_3^2-1\right) \left(s_0^2 s_1^2 s_2^2 s_3^2-1\right) }{s_0^5 s_1^3 s_2^3 \left(s_0^2 s_2^2-1\right)  s_3^3 \left(s_0^2 s_3^2-1\right)}\,,\cr
&\gamma = \frac{s_1^2 s_3^2 s_2^4+s_1^2 s_2^2-s_1^2 s_3^2 s_2^2+s_3^2 s_2^2-s_2^2+1}{s_0^2 s_2^2 s_3^2}\,,\cr
&\delta = -\frac{s_1^2}{s_0^4 s_3^2}\,,\,\,\,\,\,\, \mu = \frac{-s_2^2 s_1^2+s_2^2 s_3^2 s_1^2+s_1^2-1}{s_0 s_1 s_2 s_3}\,.
}
\e $$ 

In the case of $B_4$ models the crossing parameters are
 $$\eqalign{
 \zeta_0=7, \,\,\,\,\,\, \zeta_1=3, \,\,\,\,\,\, \zeta_2=-1, \,\,\,\,\,\, \zeta_3=-5
 }
 \e $$ 
Using the results of \r\CB\  with the above explicit parameters the desired $5$-CB algebra relations for $B_4$ models can be found.  
The $B_4$ skein relation reads explicitly
$$\eqalign{
&G_i^3=\frac{\left(q^{16}+q^{12}-q^{10}-q^6+q^4+1\right)}{q^{14}}G_i -\frac{1}{q^{12}}G_i^{-1}+\cr
&+\frac{(q^2-1)\left(q^4+1\right) \left(q^6-q^2-1\right)}{q^{10}}G_i^2 +\frac{\left(q^{12}-q^6-q^2+1\right)}{q^{14}}1_i+\cr
&+\frac{\left(q^4+1\right) \left(q^4-q^3+q^2-q+1\right) \left(q^4+q^3+q^2+q+1\right) \left(q^{12}-q^6+1\right) \left(q^2-1\right)^2}{q^{38}}E_i\;.
}
 $$
To shorten notation, we denote below by $a_{l,m,n}$ the elements of the algebra $A_{l}[i] A_{m}[i+1] A_{n}[i]$ and 
by $b_{l,m,n}$ the elements of the algebra $A_{l}[i+1] A_{m}[i] A_{n}[i+1]$,
where $A_l[r]$ stands for $G_r, G_r^{-1},E_r,G_r^{2}$ or $1_r$ according to whether $l=1,2,3,4,5$, respectively.
 A few of the relations,  which are sufficiently short, are listed below:
 $$\eqalign{
1)\,\,\,&\frac{\left(q^{12}-q^{10}-q^6+q^4+1\right) a_{5,2,3}}{q^{10}}+\frac{\left(q^{12}-q^{10}-q^6+q^4+1\right) a_{5,3,1}}{q^{10}}-\cr
&-\frac{\left(q^{12}-q^{10}-q^6+q^4+1\right) a_{5,1,3}}{q^{10}}-\frac{\left(q^{12}-q^{10}-q^6+q^4+1\right) a_{5,3,2}}{q^{10}}-\cr
&-\frac{\left(q^{10}+q^6-q^4+q^2-1\right) \left(q^{12}-q^6+1\right) a_{5,5,3}}{q^{20}}-a_{4,3,3}-a_{5,3,4}+a_{5,4,3}+\cr
&\frac{\left(q^{10}+q^6-q^4+q^2-1\right) \left(q^{12}-q^6+1\right) b_{5,5,3}}{q^{20}}+b_{3,3,4}=0\cr
}
 $$
$$\eqalign{
2)\,\,\,&\frac{\left(q^4+1\right) \left(q^{12}-q^6+1\right) a_{5,5,3}}{q^{20}}-\frac{(q^2-1)  \left(q^4+1\right) \left(q^6+q^4-1\right) b_{5,2,3}}{q^{20}}+\cr
&+\frac{(q^2-1)  \left(q^4+1\right) \left(q^6+q^4-1\right) b_{5,3,2}}{q^{20}}+\frac{b_{5,4,3}-b_{5,3,4}}{q^6}-\cr
&-\frac{\left(q^4+1\right) \left(q^{12}-q^6+1\right) b_{5,5,3}}{q^{20}}+\frac{(q^2-1)  \left(q^4+1\right) \left(q^6-q^2-1\right) b_{5,3,1}}{q^{16}}-\cr
&-\frac{(q^2-1) \left(q^4+1\right) \left(q^6-q^2-1\right) b_{5,1,3}}{q^{16}}+b_{1,4,3}-a_{3,4,1}=0
}
$$
$$\eqalign{
3)\,\,\,&\frac{\left(q^{10}+q^6-q^4+q^2-1\right) \left(q^{12}-q^6+1\right) (a_{5,4,3}-a_{5,3,4}+b_{5,3,4}-b_{5,4,3})}{q^{10} \left(q^{12}-q^{10}-q^6+q^4+1\right)}-\cr
&-\frac{\left(q^{10}+q^6-q^4+q^2-1\right)^2 \left(q^{12}-q^6+1\right)^2 (a_{5,5,3}-b_{5,5,3})}{q^{30} \left(q^{12}-q^{10}-q^6+q^4+1\right)}+\cr
&+\frac{\left(q^{10}+q^6-q^4+q^2-1\right) \left(q^{12}-q^6+1\right) (b_{5,1,3}+b_{5,3,2}-b_{5,2,3}-b_{5,3,1})}{q^{20}}+\cr
&+\frac{q^{10} (a_{4,3,4}- b_{4,3,4})}{q^{12}-q^{10}-q^6+q^4+1}-a_{1,3,4}+a_{2,3,4}-b_{4,3,2}+b_{4,3,1}=0
}
 $$
 $$\eqalign{
4)\,\,\,&\frac{(q^2-1) \left(q^4+1\right) \left(q^{12}-q^6+1\right) \left(q^{18}-q^{16}+q^{10}-q^8-1\right) (b_{5,3,1}-b_{5,1,3})}{q^{20}}+\cr
&+\frac{(q^2-1)  \left(q^4+1\right) \left(q^6-q^2-1\right) (b_{5,2,3}-b_{5,3,2})}{q^4}+b_{2,4,3}+\cr
&+\frac{\left(q^4+1\right) \left(q^{12}-q^6+1\right) \left(q^{12}-q^{10}+q^8-q^2+1\right) (a_{5,5,3}-b_{5,5,3})}{q^6}++\cr
&+\left(q^{12}-q^{10}-q^6+q^4+1\right) q^2 (b_{5,4,3}-b_{5,3,4})-a_{3,4,2}=0\cr
}
 $$
  $$\eqalign{
5)\,\,\,&\frac{(q^2-1) \left(q^4+1\right) \left(q^{12}-q^6+1\right) \left(q^{18}-q^{16}+q^{10}-q^8-1\right) (a_{5,1,3}-a_{5,3,1})}{q^{20}}+\cr
&+\frac{\left(q^4+1\right) \left(q^{12}-q^6+1\right) \left(q^{12}-q^{10}+q^8-q^2+1\right) (a_{5,5,3}- b_{5,5,3})}{q^6}+\cr
&+\frac{(q^2-1)  \left(q^4+1\right) \left(q^6-q^2-1\right) (a_{5,3,2}-a_{5,2,3})}{q^4}+\cr
&+\left(q^{12}-q^{10}-q^6+q^4+1\right) q^2 (a_{5,3,4}- a_{5,4,3})-a_{2,4,3}+b_{3,4,2}=0
}
 $$
  $$\eqalign{
6)\,\,\,&\frac{\left(q^{12}-q^{10}-q^6+q^4+1\right) (a_{1,3,1}-b_{1,3,1})}{q^{10}}-a_{2,3,4}-a_{4,3,1}+b_{1,3,4}+b_{4,3,2}+\cr
&+\frac{\left(q^{10}+q^6-q^4+q^2-1\right) \left(q^{12}-q^6+1\right) (a_{5,3,1}-a_{5,1,3}+b_{5,2,3} - b_{5,3,2})}{q^{20}}=0
}
 $$
  $$\eqalign{
7)\,\,\,&\frac{\left(q^8+1\right) \left(q^8-q^4+1\right) \left(q^{16}+1\right) \left(q^4+1\right)^2 (a_{5,5,3}-b_{5,5,3})}{q^{20}}-a_{3,4,3}+b_{3,4,3}=0
}
 $$
  $$\eqalign{
8)\,\,\,\,\,\,\,\,\,&\frac{a_{5,3,4}-a_{5,4,3}}{q^6}+\frac{(q^2-1) \left(q^4+1\right) \left(q^6+q^4-1\right) (a_{5,2,3}-a_{5,3,2})}{q^{20}}-a_{1,4,3}+\cr
&+\frac{\left(q^4+1\right) \left(q^{12}-q^6+1\right) (a_{5,5,3}-b_{5,5,3})}{q^{20}}+b_{3,4,1}+\cr
&+\frac{(q^2-1) \left(q^4+1\right) \left(q^6-q^2-1\right) (a_{5,1,3}- a_{5,3,1})}{q^{16}}=0
}
 $$

We find that the whole list of 19 5-CB relations, which can be found in the attached Mathematica file, is fulfilled for the  Boltzmann weights of $B_4$ models. The Boltzmann weights are stated in \r\Bk.  We checked the relations numerically for a general value of the parameter $q$ and substituting  various heights.

\mysec{4--CB relations for $D_6$.}

We wish to check the 4--CB algebra for $D_6$ which is a four blocks model.
The four blocks relations  were given in \r\CBfour. 
The parameters for $D_6$ are, eq. (2.18).
$$\zeta_0=10,\qquad \zeta_1=6,\qquad \zeta_2=2,\e$$
and $q$ is given by eq. (2.7).

The skein relation is \r\CBfour,
\def\frac#1#2{{#1\over #2}}
$$\threeline{
G_i^2= -i q^{-\frac{1}{2}  \zeta_0-\zeta_1-\zeta_2} \left(1-q^{2 \zeta_1}+q^{2 \zeta_1+2  \zeta_2}\right) \
G_i-i q^{-\frac{3}{2}  \zeta_0+\zeta_1-\zeta_2} \
G_i^{-1}}{+\frac{q^{-2  \zeta_0-2  \zeta_1-2 \zeta_2} \left(q^{2  
\zeta_1}-1\right) \left(1+q^{2  \zeta_0+2 
\zeta_1+2 \zeta_2}\right) \left(q^{2  \zeta_2}-1\right) 
 }{\left(q^{2  \zeta_0+ 2  \zeta_2}-1\right) }E_i}{-q^{-\zeta_0-2  
\zeta_2} \left(1-q^{2  \zeta_2}+q^{2  \zeta_1+2 \zeta_2}\right).}$$

The single additional relation is 
$$g(i,i+1,i)=g(i+1,i,i+1),\e$$
where 
$$\threeline{
g=a_{1,2,4}+a_{1,3,1}+a_{4,2,1}-i q^{-\zeta_0/2+\zeta_1-\zeta_2} (a_{1,3,4}+a_{4,2,4}+a_{4,3,1})-}{
i q^{\zeta_0/2-\zeta_1+\zeta_2}  (a_{2,3,4}+a_{4,1,4}+a_{4,3,2})-}{
i {q^{\zeta_1+\zeta_2}\over (q^{2\zeta_1}-1)(q^{2\zeta_2}-1)}\left(q^{\zeta_0/2} a_{1,2,1}+q^{-\zeta_0/2} a_{2,1,2} \right)+z a_{4,3,4},}$$
where 
$$\twoline{z={q^{-\zeta_0-2\zeta_1-2\zeta_2} (q^{2\zeta_1}-1)(q^{2\zeta_2}-1)\over q^{2\zeta_0+2\zeta_2}-1}\times}{
\left( 2 q^{2\zeta_0+2\zeta_2} +2 q^{2\zeta_0+2\zeta_1+2\zeta_2}+q^{4\zeta_0+2\zeta_1+4 \zeta_2}+1\right).}$$  
We denoted by $a_{i,j,k} (r,s,t)$ the element of the algebra $a_i[r] a_j[s] a_k[t]$ where
$a_i[r]$ is $G_r, G_r^{-1},E_r$ or $1_r$, if $i=1,2,3,4$, respectively. 

Finally, we proceed to check these two relations, for the $D_6$ vertex model substituting the explicit Boltzmann weights, eqs. (2.20, 2.21).
Indeed they hold for various values of the heights and for general value of $q$.

\ack
We thank Ida Deichaite for remarks on the manuscript.

                                                                                                                                                                                                                                                                                                                                                                                                                                                                                                    

\refout

\bye